\documentclass[amsmath,amssymb,aps,prl,twocolumn]{revtex4-1}	% arXiv
\usepackage{txfonts}
\usepackage{graphicx}
\usepackage{bm}
\usepackage{array}
\usepackage{amsmath}

\newcommand{\ET}{$\kappa$-(BEDT-TTF)$_2$Cu$_2$(CN)$_3$}
\newcommand{\dmit}{EtMe$_3$Sb[Pd(dmit)$_2$]$_2$}
\newcommand{\BCSO}{Ba$_3$CuSb$_2$O$_9$}
\newcommand{\dmitCO}{Et$_2$Me$_2$Sb[Pd(dmit)$_2$]$_2$}
\newcommand{\Hcat}{$\kappa$-H$_3$(Cat-EDT-TTF)$_2$}

\begin{document}

\title{Boundary-limited and glassy-like phonon thermal conduction in EtMe$_3$Sb[Pd(dmit)$_2$]$_2$}

\author{Minoru Yamashita}

\affiliation{
The Institute for Solid State Physics, The University of Tokyo, Kashiwa, 277-8581, Japan
}

%\abst{
\begin{abstract}
In  molecular-based quantum-spin-liquid candidate EtMe$_3$Sb[Pd(dmit)$_2$]$_2$ with two-dimensional $S$=1/2 triangular lattice, a finite  residual linear term in the thermal conductivity,  $\kappa_0/T\equiv\kappa/T (T \rightarrow 0)$,  has been observed and attributed to the presence of itinerant gapless excitations.  
Here we show that the data of $\kappa$ measured in several single crystals are divided into two groups with and without the residual linear term.  In the first group with finite $\kappa_0/T$,  the phonon thermal conductivity $\kappa_{ph}$ is comparable to that of other organic compounds.  In these crystals, the phonon mean free path $\ell_{ph}$ saturates at low temperatures, being limited by the sample size.
On the other hand, in the second group with zero $\kappa_0/T$,  $\kappa_{ph}$ is one order of magnitude smaller than that in the first group, comparable to that of amorphous solids.
In contrast to the first group, $\ell_{ph}$ shows a glassy-like non-saturating behavior at low temperatures.
These results suggest that the crystals with long $\ell_{ph}$ are required to discuss  the magnetic excitations by thermal conductivity measurements.
\end{abstract}

\maketitle

Quantum spin liquids (QSLs)~\cite{Balents2010} are novel state of matter, in which  the strong quantum fluctuations melt the magnetic order even at zero temperature. 
The ground states of QSLs have attracted much attention for decades because of the emergence of exotic elementary excitations, such as  spinons in the one-dimension (1D) QSL and itinerant Majorana fermions in the Kitaev QSL~\cite{Kitaev2006}.  QSLs  are frequently found in a class of materials known as frustrated magnets. 
Candidate materials hosting the QSLs have been found in various materials with triangular \cite{KanodaKato2011, Isono2014,LawPALee2017, YueshengLi_2015_SciRep, YueshengLi_2015_PRL}, kagome \cite{ImaiLee2016}, honeycomb \cite{Nakatsuji2012} and pyrochlore \cite{Gardner1999} lattices.

To reveal  the nature of QSL states, it is crucially important to clarify whether the low-lying excitations are gapped or gapless,  and whether they are localized or itinerant.
The specific heat ($C$) and thermal conductivity ($\kappa$) measurements provide vital information on these issues.  The former includes both localized and itinerant excitations, while the latter sensitively detects the itinerant low-lying excitations, which is not contaminated by localized impurities.  
Gapless itinerant excitations have been reported by thermal conductivity measurements in some of QSL candidate materials including organic {\dmit}\cite{MYama2010} and $\kappa$-H$_3$(Cat-EDT-TTF)$_2$~\cite{Shimozawa2017}, and inorganic Tb$_2$Ti$_2$O$_7$~\cite{HirschbergerScience2015} and 1T-TaS$_2$~\cite{Murayama_arXiv180306100}.
In these materials,  the gapless excitations have been discussed in terms of  spinon Fermi surface \cite{Motrunich2006,SLee2005}.
On the other hand, the absence of the gapless excitations has been reported in {\ET} \cite{MYamashita2008} and YbMgGaO$_4$ \cite{YXu2016}, which has been discussed in terms of a gapped QSL \cite{Pratt2011}, inhomogeneity effects \cite{MYamashita2012}, and scattering effect on spin excitations by disorder \cite{YueshengLi2017}.

In {\dmit} \cite{KanodaKato2011}, dimerized Pd(dmit)$_2$ molecules form 2D triangular lattice of spin $S$ = 1/2 which is separated by layers of non-magnetic cation EtMe$_3$Sb$^+$.
Despite  the large  exchange energy of  $J\sim 235$ K \cite{Itou2008}, no evidence of  magnetic order has been reported  down to $\sim19$ mK ($\sim J / 12000$) \cite{Itou2010}.
The presence of gapless excitations has been reported  by the specific heat  \cite{SYamashita2011} and magnetic susceptibility \cite{Watanabe2012} measurements.  Moreover, the thermal conductivity measurements have  revealed  the presence of a finite  residual linear term,  $\kappa_0/T\equiv\kappa/T (T \rightarrow 0)$, indicating   that  the gapless excitations contain itinerant contributions~\cite{MYama2010}.

Recently, the absence of  $\kappa_0/T$ in  {\dmit}  has been reported by two groups \cite{BourgeoisHope_arXiv190410402,SYLi_arXiv190410395}. 
In this letter, we reinvestigate the thermal conductivity of {\dmit}.  We show that there are two types of crystals with zero and finite $\kappa_0/T$.   We find that the phonon thermal conductivity $\kappa_{ph}$ of crystals with finite $\kappa_0/T$, which is comparable to $\kappa_{ph}$ of other organic compounds, is much larger than  $\kappa_{ph}$ of crystals with zero $\kappa_0/T$.  

\begin{figure}[!tbh]
\centering
\includegraphics[width=\linewidth]{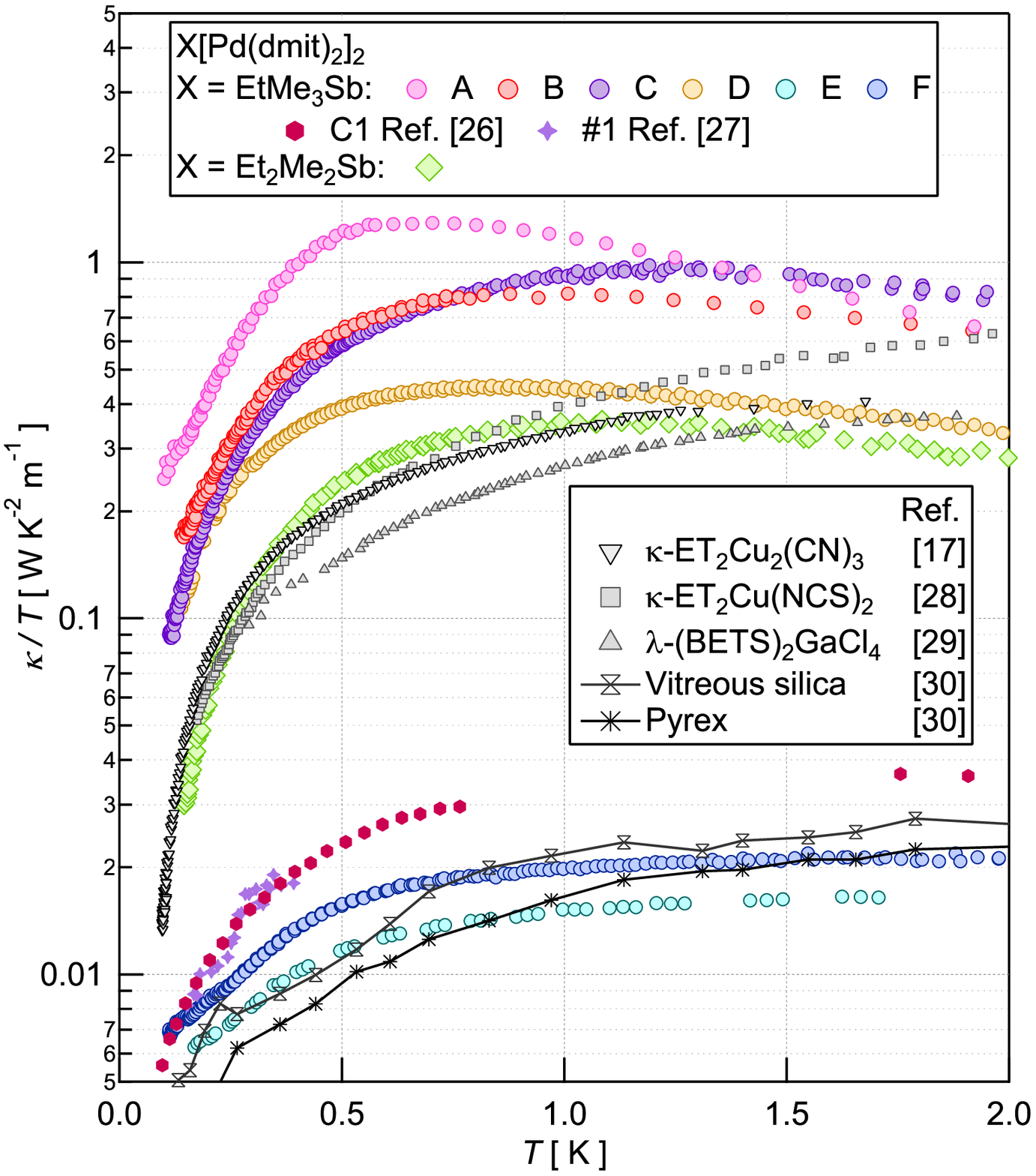}
\caption{Temperature dependence of thermal conductivity divided by temperature ($\kappa / T$) of the QSL candidate {\dmit} and the non-magnetic compound Et$_2$Me$_2$Sb[Pd(dmit)$_2$]$_2$. The data of sample A and B of EtMe$_3$Sb and that of Et$_2$Me$_2$Sb are the same data in ref. \cite{MYama2010}.
The data of {\dmit} by other groups (sample C1 from ref. \cite{BourgeoisHope_arXiv190410402} and sample \#1 from ref. \cite{SYLi_arXiv190410395}) are also plotted.
Thermal conductivity of other organic compounds (a QSL candidate with two-dimensional triangular lattice {\ET} \cite{MYamashita2008} and quasi two-dimensional superconductors $\kappa$-(BEDT-TTF)$_2$Cu(NCS)$_2$ \cite{Belin1998} and $\lambda$-(BETS)$_2$GaCl$_4$ \cite{Tanatar2002})  and that of amorphous solids (vitreous silica and pyrex~\cite{ZellerPohl1971}) are shown together for comparison.}
\label{fig1}
\end{figure}

Figure\,1 summarizes the temperature dependence of $\kappa / T$ of {\dmit} obtained  from different batches, along with that of other materials. Thermal conductivity was measured by the standard steady-state method \cite{MYama2010,MYamashita2008}.
The samples were cooled down slowly with the rate of -10--30 K/hour and -100 K/hour for sample A-E and for sample F, respectively. For comparison, we also plot $\kappa / T$ of {\dmit} from refs. \cite{BourgeoisHope_arXiv190410402, SYLi_arXiv190410395}, the non-magnetic compound {\dmitCO} \cite{MYama2010}, another QSL candidate {\ET} \cite{MYamashita2008}, quasi two-dimensional superconductors $\kappa$-(BEDT-TTF)$_2$Cu(NCS)$_2$ \cite{Belin1998} and $\lambda$-(BETS)$_2$GaCl$_4$ \cite{Tanatar2002}, and amorphous solids (vitreous silica and pyrex) \cite{ZellerPohl1971}.
It is obvious that $\kappa/T$  of {\dmit} are divided into two groups.  One group including samples A, B,C and D has a large $\kappa / T$ value, which is comparable to  other organic compounds.  We refer this group as large-$\kappa$ group.  On the other hand,  the other group including samples E and F has a small $\kappa/T$ value, which is one order of magnitude smaller than that of the large-$\kappa$ group and is  comparable to amorphous solids \cite{ZellerPohl1971}.  We refer this group as small-$\kappa$ group.  We note that the magnitude and temperature dependence of $\kappa/T$ in the small-$\kappa$ group are similar  to  those  reported in refs. \cite{BourgeoisHope_arXiv190410402, SYLi_arXiv190410395}.

Thermal conductivity of magnetic insulators consists of phonon  and  spin contributions.
Since phonon contribution  is usually dominant above 1\,K,   the observed large difference in $\kappa/T$ points to very different  $\kappa_{ph}$ between the two groups  of {\dmit}. It has been reported that the specific heat of {\dmit} with large $\kappa/T$ is close to that with small  $\kappa/T$ \cite{SYamashita2011, SYLi_arXiv190410395}.  As  $\kappa_{ph}$ is given by $\frac{1}{3} C_{ph} v_{ph} \ell_{ph}$, where $C_{ph}$ is the specific heat of phonons, $v_{ph}$ the sound velocity, and $\ell_{ph}$ the mean free path of phonons, the large difference of $\kappa/T$ is attributed to the lager difference of $\ell_{ph}$.

\begin{figure}[!tbh]
\centering
\includegraphics[width=0.9 \linewidth]{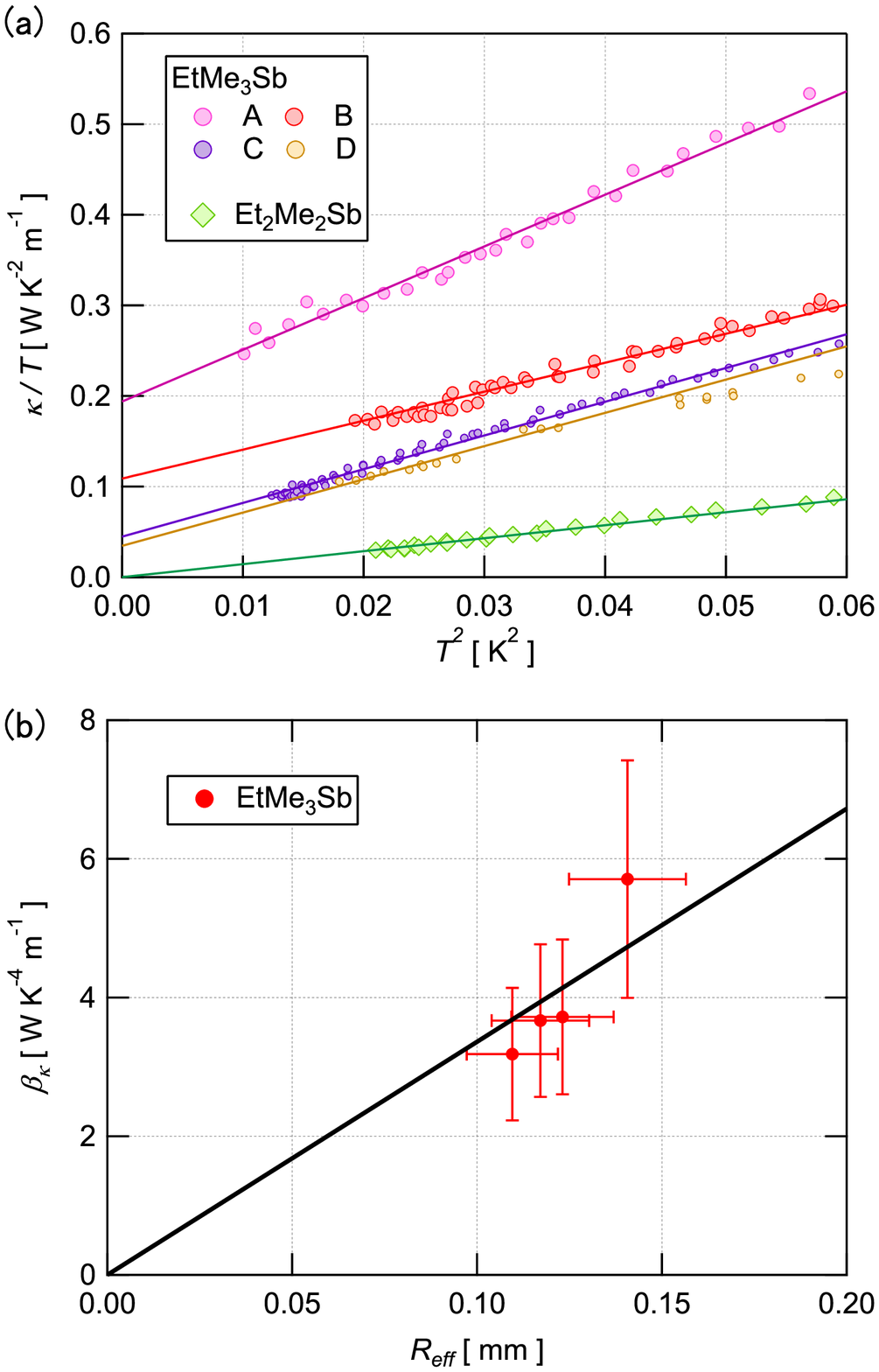}
\caption{Low temperature data of $\kappa / T$ of {\dmit} in the large-$\kappa$ group as a function of $T^2$.
All data of {\dmit} show $\kappa / T = \kappa_0 / T + \beta_{\kappa} T^2$, whereas
the non-magnetic Et$_2$Me$_2$Sb[Pd(dmit)$_2$]$_2$ shows only the phonon contribution.
(b) The coefficient $\beta_{\kappa}$ and the effective sample dimension $R_{eff}$ for the samples of A--D.
The error estimating the sample dimensions causes the error bar of the data.
The solid line shows the linear fit of the data.
See the main text for detail.
}
\label{fig2}
\end{figure}

Figure\,2 (a) depicts $\kappa / T$ vs.\,$T^2$ in the low temperature regime of {\dmit} of the large-$\kappa$ group, along with non-magnetic Et$_2$Me$_2$Sb[Pd(dmit)$_2$]$_2$. 
These {\dmit} crystals show the temperature dependence of $\kappa / T = \kappa_0 / T + \beta_{\kappa} T^2$ with a finite $\kappa_0 / T$.
In the non-magnetic Et$_2$Me$_2$Sb[Pd(dmit)$_2$]$_2$,  $\kappa_0 / T$ is absent. The observed $T^2$-dependence of $\kappa/T$ is a typical behavior of phonon conduction in the boundary scattering limit, where $\ell_{ph}$  is limited by the sample size.   This can be quantitatively supported by the following estimation.  
In the boundary scattering regime, the effective sample size is  given by $R_{eff} \approx  \sqrt{w \cdot t}$,  where $w$ and $t$ is the sample width and the thickness, respectively.  Figure\,2(b) depicts $\beta_{\kappa}$, which is obtained by the slope of $\kappa/T$ vs. $T^2$ in Fig.\,2(a), plotted as a function of $R_{eff}$ for samples A-D.   $\beta_{\kappa}$ increases linearly with $R_{eff}$   within the error range.    We estimate the sound velocity from the linear relationship shown by the solid line  in Fig.\,2(b).  The sound velocity is  estimated by $v_{ph}=3(\beta_{\kappa}/R_{eff})/\beta_C$, where $\beta_{C}$ is the coefficient of $T^3$-term in the specific heat.  By using  $\beta_{C} = 22.8$ mJ K$^{-4}$ mol$^{-1}$ from the reported  specific heat data \cite{SYamashita2011}, we  obtain  $v_{ph} \sim 2100$ m/s.   This value is comparable to $v_{ph} \sim 1400$\,m/s  estimated from the Debye relation, $\beta_C = \frac{2 \pi^2}{5} k_B ( k_B / \hbar v_{ph})^3$.
These results  indicate that  $\kappa_{ph}$ of {\dmit} in the large-$\kappa$ group is in the boundary scattering limit at low temperatures.

\begin{figure}[!tbh]
\centering
\includegraphics[width=\linewidth]{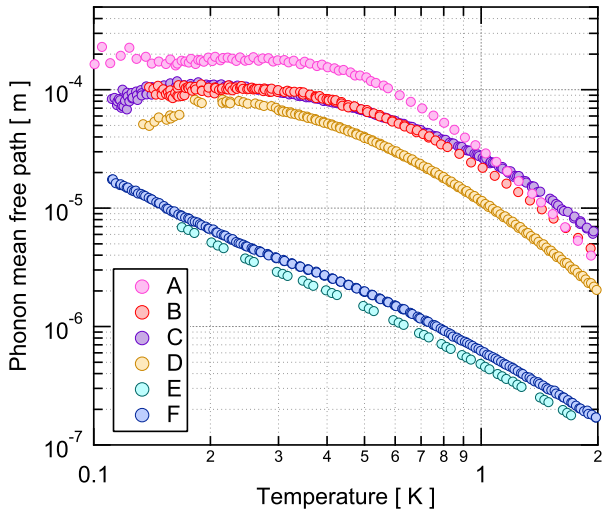}
\caption{The temperature dependence of the phonon mean free path $\ell_{ph}$ of {\dmit}. See the main text for detail.
}
\label{fig3}
\end{figure}

Figure\,3 depicts the temperature dependence of  $\ell_{ph}$ for  {\dmit} below 2\,K.  In this temperature range, specific heat shows  $T^3$-dependence \cite{SYamashita2011}, indicating $C_{ph}=\beta_C T^3$.   We evaluate $\ell_{ph}$ by  the relation $\kappa_{ph} = \beta_{C} T^3 v_{ph} \ell_{ph}/3$ using  $v_{ph}=2100$\,m/s.  
As shown in Fig.\,3, $\ell_{ph}$ in the large-$\kappa$ group saturates below $\sim 0.5$ K due to the boundary scattering.
In sharp contrast, $\ell_{ph}$ in the small-$\kappa$ group increases as the temperature is lowered without exhibiting saturating behavior. 
We point out that this non-saturating behavior of $\ell_{ph}$ bears resemblance to that observed  in amorphous solids \cite{Berman}.
In fact, as shown in Fig.\,1, $\kappa$ in the small-$\kappa$ group is similar to that of vitreous silica or pyrex \cite{ZellerPohl1971}.
We note that the glassy-like thermal conductions have been observed even in crystalline materials such as clathrate compounds \cite{Takabatake2014}, Tb$_2$Ti$_2$O$_7$ \cite{QJLi2013}, and {\BCSO} \cite{Sugii2017}.
In these materials, $\kappa_{ph}$ is suppressed by a rattling of the guest atoms \cite{Takabatake2014}, a strong spin fluctuations \cite{QJLi2013}, and random domains \cite{Sugii2017}.

We discuss here several possible origins for the large difference of $\kappa_{ph}$  between the two groups of {\dmit}.  
First is the influence of phonon scattering by spin excitations suggested in refs. \cite{BourgeoisHope_arXiv190410402,SYLi_arXiv190410395}.  Similar effects have also been discussed in Tb$_2$Ti$_2$O$_7$ \cite{QJLi2013} and YbMgGaO$_4$ \cite{YXu2016}.  
However, the large different spin-phonon scattering rate between the two groups  consisting of the same molecules  is unlikely.   
Second is the structural domain formation. 
In {\dmit}, the non-centrosymmetric cations EtMe$_3$Sb$^+$ have two orientations in the crystal \cite{TamuraKato2009}, which may give rise to large number of domains of different sizes.   In  {\BCSO} \cite{Sugii2017}, for instance, the Cu$^{2+}$-Sb$^{5+}$ dumbbells have the Ising degree of freedom in the structure, giving rise to domains of random size structures \cite{Wakabayashi2016, Smerald2015}.  Third is the microcracks.  In thermal conductivity measurements, unavoidable mechanical stress is applied on the crystal, which often leads to the formation of microcracks in organic compounds.  {\dmit} may be sensitive to such a stress.  
In all cases, the specific heat measurements cannot distinguish between the large and small $\kappa$ groups, whereas $\kappa$ exhibits remarkably different behavior between the two groups.  
In the second and the third cases, $\ell_{ph}$ is determined by the domains with  broad size distribution, giving rise to a strong suppression of $\kappa_{ph}$ and  non-saturating temperature dependence of $\ell_{ph}$, similar to amorphous solids.
Further studies are necessary to resolve these issues.

Finite $\kappa_0/T$ is observable only in {\dmit} of the large-$\kappa$ group.   As shown in Fig.\,2(a),  the magnitude of $\kappa_0/T$ is strongly sample dependent, implying that the mean free path of the spin excitations are extremely sensitive to the impurities.  Therefore, the absence of $\kappa_0/T$ in the small-$\kappa$ group may imply that {\dmit} of small-$\kappa$ group contains higher concentration of impurities.     
We also note that thermal conductivity studies of other organic compounds with large $\kappa_{ph}$ have successfully detected the magnetic contribution.
In {\ET}, as shown in Fig. 1, $\kappa_{ph}$ is comparable to that of {\dmit} in the large-$\kappa$ group. 
At low temperatures, $\kappa$ shows an activated-temperature dependence, suggesting  a gap formation in the magnetic excitations \cite{MYamashita2008}.
Moreover, a finite $\kappa_0/T$ has been observed in {\Hcat},  in which $\ell_{ph}$  exceeds the effective sample size at low temperatures \cite{Shimozawa2017}.
These results appear to indicate that samples with long $\ell_{ph}$ and with very low impurity concentration  are crucial to study the itinerant magnetic excitations by thermal conductivity measurements.

\begin{acknowledgments}
The author thanks Yuichi Kasahara, Reizo Kato, Yuji Matsuda, and Takasada Shibauchi for fruitful discussions. 
\end{acknowledgments}

%\bibliography{dmit_k}

\end{document}